# Proposal for Quantum Sensing Based on Two-dimensional Dynamical Decoupling: NMR Correlation Spectroscopy of Single Molecules

Wen-Long Ma and Ren-Bao Liu[*]

*Department of Physics and Centre for Quantum Coherence, The Chinese University of Hong Kong, Shatin, New Territories, Hong Kong, China*

**Abstract**

Nuclear magnetic resonance (NMR) has enormous applications. Two-dimensional NMR has been an essential technique to characterize correlations between nuclei and hence molecule structures. Towards the ultimate goal of single-molecule NMR, dynamical-decoupling- (DD) enhanced diamond quantum sensing has enabled detection of single nuclear spins and nanoscale NMR. However, there is still lack of a standard method in DD-based quantum sensing to characterize correlations between nuclear spins in single molecules. Here we present a scheme of two-dimensional DD-based quantum sensing, as a universal method for correlation spectroscopy of single molecules. We design two-dimensional DD sequences composed of two sets of periodic DD sequences with different periods, which can be independently set to match two different transition frequencies for resonant DD. We find that under the resonant DD condition the sensor coherence patterns, as functions of the two independent pulse numbers of DD subsequences, can fully determine different types of correlations between nuclear spin transitions. This work offers a systematic approach to correlation spectroscopy for single-molecule NMR.

---

[*] Corresponding author. Email: rbliu@phy.cuhk.edu.hk

# I. Introduction

Nuclear magnetic resonance (NMR) has important applications in analytical chemistry, structural biology and quantum computing [1-3]. Furthermore, two-dimensional (2D) NMR [4], which employs composite control pulses and sequences on nuclear spins, provides information about the couplings between the nuclear spins and hence the molecule structures. Two-dimensional spectroscopy has also been extended to optics to study correlations in molecules [5] and many-body effects in semiconductors [6-7]. However, the conventional NMR is generally performed for large ensembles of molecules ($>10^{12}$ molecules) and at high magnetic fields (>1 Tesla). The top challenge of magnetic spectroscopy is NMR with atomic-scale sensitivity and resolution [2].

A breakthrough toward single-spin NMR is dynamical-decoupling- (DD) based quantum sensing [8, 9]. The basic principle is as follows. The quantum sensor (such as the electron spin of a nitrogen-vacancy (NV) center in diamond [10]) loses its coherence due to noises from weakly coupled nuclear spins (targets). The DD control [Fig. 1(a)] flips the sensor state periodically so as to cancel the effect of the background noises. The noises from the target nuclear spins have characteristic frequencies corresponding to their transitions. If the DD period matches the characteristic frequency, the noise from the target nuclear spins is resonantly amplified and the sensor coherence presents fingerprint sharp dips. Using this DD-based scheme, several groups have successfully detected single $^{13}$C nuclear spins [11-13] and $^{13}$C clusters in diamond [14]. Shallow NV centers near diamond surfaces have been used to sense NMR of single nuclear spins [15], nano-scale NMR of molecules [16-17] and single protein molecules [18]. This DD-based quantum sensing, however, has important limitations: (i) it does not resolve nuclear spins of the same species; (ii) it does not distinguish nuclear spin clusters of different correlation types if the clusters produce the same noise spectrum. It is highly desirable to have a correlation spectroscopy in quantum sensing similar to that in 2D NMR [19]. Recently we have found that for DD resonant with a specific target spin transition, the sensor coherence dip oscillates periodically as a function of DD pulse number, with the oscillation period inversely proportional to the coupling strength between the sensor spin and target spin [20]. Based on this finding, we proposed a scheme that is capable of resolving single nuclear spins of the same species and identifying correlations in nuclear spin clusters [20]. However, that scheme cannot fully differentiate different types of correlations in nuclear spin clusters.

In this work, we propose a scheme of 2D DD-based quantum sensing, which is capable of fully characterizing the correlations in single nuclear spin clusters. The 2D DD sequence is composed of two sets of periodic DD sub-sequences with different pulse intervals. When the pulse intervals match different nuclear spin transitions (resonant DD), the sensor coherence, as a function of the pulse numbers of the DD sub-sequences, shows distinct patterns depending on the correlation types of different nuclear spin transitions.

## II. Theoretical model

We consider a spin-1/2 quantum sensor ($S=1/2$) weakly coupled to $M$ target clusters of nuclear spins (each cluster representing a "molecule"). The general Hamiltonian is

$$H = S_z \sum_{k=1}^{M} \beta^{(k)} + \sum_{k=1}^{M} H_0^{(k)}, \qquad (1)$$

where $H_0^{(k)} = \sum_{m=1}^{d_k} \varepsilon_m^{(k)} |m\rangle_k {}_k\langle m|$ is the nuclear spin Hamiltonian for the $k$-th "molecule" with $d_k$ denoting the number of eigenstates $\{|m\rangle_k\}$, $\beta^{(k)} = \frac{1}{2} \sum_{m,n} \left( \beta_{mn}^{(k)} |m\rangle_k {}_k\langle n| + \text{H.c.} \right)$ is the noise operator from the $k$-th "molecule" which induces the nuclear spin transition $|m\rangle_k \leftrightarrow |n\rangle_k$ with the transition frequency $\omega_{mn}^{(k)} = \varepsilon_m^{(k)} - \varepsilon_n^{(k)}$ and the transition matrix element $\beta_{mn}^{(k)} = {}_k\langle m|\beta^{(k)}|n\rangle_k$. In quantum sensing, the coupling to the target spins is weak, i.e. $|\beta_{mn}^{(k)}| \ll |\omega_{mn}^{(k)}|$ for $m \neq n$. We denote the eigenstates of the sensor operator $S_z$ as $|\pm\rangle$.

To suppress the background noise [21] and selectively enhance the noise from the target spins [11], we apply DD [8, 9] to the sensor (consisting of a sequence of $\pi$-flips at times $\{t_1, t_2 \cdots t_N\}$ for the evolution from 0 to $t$). Under DD control, the target spin evolution conditioned on the sensor states $|\pm\rangle$ is $U_N^{(\pm)}(t) = \bigotimes_k U_{k,N}^{(\pm)}(t)$ with

$$U_{k,N}^{(\pm)}(t) = e^{-i[H_0^{(k)} \pm (-1)^N \beta^{(k)}/2](t-t_N)} \cdots e^{-i(H_0^{(k)} \mp \beta^{(k)}/2)(t_2-t_1)} e^{-i(H_0^{(k)} \pm \beta^{(k)}/2)t_1} . \qquad (2)$$

The sensor spin decoherence is caused by the bifurcated quantum evolution of the target spins conditioned on the sensor state,

$$L(t) = \prod_{k=1}^{M} \text{Tr}_k \left[ \rho_k \left( U_{k,N}^{(-)} \right)^\dagger U_{k,N}^{(+)} \right], \qquad (3)$$

where the density matrix of the $k$-th target is assumed in a maximally mixed state $\rho_k = d_k^{-1} \sum_{m=1}^{d_k} |m\rangle_{kk}\langle m|$ as the temperature is usually much higher than the nuclear spin transition frequencies. In the following, we will omit the subscript $k$ when we consider only one target "molecule".

## III. One-dimensional DD-based quantum sensing

One-dimensional (1D) DD sequence is just the conventional $N_1$-pulse Carr-Purcell-Meiboom-Gill (CPMG-$N_1$) sequence [22, 23]. The pulse interval is $2\tau_1$ and $t_{p_1} = (2p_1 - 1)\tau_1$ with $p_1 = 1, 2, \cdots, N_1$, as shown in Fig. 1(a). As shown in Fig. 1(c), the sensor coherence presents dips when the resonant DD condition is realized, i.e., when the pulse interval matches the frequency of a transition $|m\rangle \leftrightarrow |n\rangle$, $2\tau_1 = \pi(2c - 1)/\omega_{mn}$ with $c = 1, 2, \cdots$. In the following we always consider the first-order coherence dip ($c = 1$). Under the resonant DD condition, the sensor spin coherence dip periodically oscillates as a function of the CPMG pulse number $N_1$ [20] (See Appendix A), namely,

$$L^{\text{dip}}(d, \delta_1, N_1) \approx \frac{1}{d} \text{Tr}\left[ U_{mn}(2N_1) \right] = \frac{1}{d}\left[ d - 2 + 2\cos(2N_1\delta_1) \right], \qquad (4)$$

where $\delta_1 = |\beta_{mn}/\omega_{mn}|$ and $U_{mn}(N_1) = \exp\left[ -iN_1 \left( \beta_{mn} |m\rangle\langle n| + \text{H.c.} \right)/\omega_{mn} \right]$. The sensor coherence has quantized minima determined by the Hilbert space dimension $d$ of the target nuclear spin cluster [ $\min\left( L^{\text{dip}}(N_1) \right) = (d-4)/d$ at $N_1 = \pi/(2\delta_1)$ ]. The quantized minima can be understood from the probability amplitude for the target spin cluster

remaining in its initial state after the evolution. The DD control on the central spin can be understood, to the leading order, as an effective AC magnetic field on the target nuclear spins, with its sign conditioned on the central spin state [20]. If the target spin cluster is initially in the state $|m\rangle$ or $|n\rangle$, the effective AC magnetic field drives the target spin cluster to periodically oscillate between $|m\rangle$ and $|n\rangle$, while if the target spin cluster is initially in other states, the DD control would leave the target spin states unchanged. Thus the depth of the sensor coherence oscillation is $-2/d + (d-2)/d = (d-4)/d$.

## IV. 2D DD-based quantum sensing

The 2D DD sequence contains two consecutive sets of CPMG sequences with pulse intervals $2\tau_1, 2\tau_2$ and pulse numbers $N_1, N_2$, respectively [Fig. 1(b)]. The sensor spin is flipped at time $t_{p_1} = (2p_1 - 1)\tau_1$, $t_{N_1+p_2} = 2N_1\tau_1 + (2p_2 - 1)\tau_2$ with $p_i = 1, 2, \cdots, N_i$ ($i = 1, 2$). The sensor coherence shows sharp dips when the two pulse intervals match two different target spin transition frequencies, i.e. $2\tau_i = \pi/\omega_i$, as shown in Fig. 1(d). The sensor coherence as a function of $N_1$ and $N_2$ contains information about the correlation between the two different transitions. The two consecutive CPMG control sequences induce two effective AC magnetic fields [with frequencies $\omega_1 = \pi/(2\tau_1)$ and $\omega_2 = \pi/(2\tau_2)$] successively applied to the target nuclear spins.

We first consider the case that the sensor spin is weakly coupled to two independent target nuclear spin "molecules" ($M = 2$). We choose a transition $|m\rangle_1 \leftrightarrow |n\rangle_1$ from target 1 and another transition $|p\rangle_2 \leftrightarrow |q\rangle_2$ from target 2. Under the double-resonance condition ($2\tau_1 = \pi/\omega_{mn}^{(1)}$, $2\tau_2 = \pi/\omega_{pq}^{(2)}$), the sub-sequences CPMG-$N_1$ and CPMG-$N_2$ resonantly amplify the noises from the transitions $|m\rangle_1 \leftrightarrow |n\rangle_1$ and $|p\rangle_2 \leftrightarrow |q\rangle_2$, respectively. Then the sensor spin coherence can be written as a product of two coherence functions $L_{1,2}^{\text{dip}}(N_1, N_2) \approx L_1^{\text{dip}}(d_1, \delta_1, N_1) L_2^{\text{dip}}(d_2, \delta_2, N_2)$ for 1D sensing as in Eq. (4) (See Appendix B), where $\delta_1 = \left|\beta_{mn}^{(1)}/\omega_{mn}^{(1)}\right|$ and $\delta_2 = \left|\beta_{mn}^{(2)}/\omega_{mn}^{(2)}\right|$ (Fig. 2).

Now we consider the case that the two different transitions $|m\rangle \leftrightarrow |n\rangle$ and $|p\rangle \leftrightarrow |q\rangle$ are in the same target "molecule". Under the double-resonance condition ($2\tau_1 = \pi/\omega_{mn}$, $2\tau_2 = \pi/\omega_{pq}$), theسensor coherence is (See Appendix C)

$$L^{\text{dip}}(N_1, N_2) \approx \frac{1}{d}\text{Tr}\left[U_{pq}(2N_2)U_{mn}(2N_1)\right] = \frac{1}{d}\sum_{a=1}^{d}\langle a|U_{pq}(2N_2)U_{mn}(2N_1)|a\rangle. \quad (5)$$

Similar to the 1D DD case, the sensor coherence can be understood as the mean probability amplitude for the target spin cluster to remain in an initial state after two consecutive evolutions $U_{mn}(2N_1)$ and $U_{pq}(2N_2)$. First let us assume the two transitions are uncorrelated (they do not share a state): (i) if the cluster in initially in state $|m\rangle$ or $|n\rangle$, the propagator $U_{mn}(2N_1)$ drives the cluster to periodically oscillate between $|m\rangle$ and $|n\rangle$ while $U_{pq}(2N_2)$ keep the cluster state unchanged, so the probability amplitude for the target spin to return to its initial state is $\cos(2N_1\delta_1)$; (ii) if the cluster is initially in state $|p\rangle$ or $|q\rangle$, the propagator $U_{pq}(2N_2)$ drives the cluster to periodically oscillate between $|p\rangle$ and $|q\rangle$ while $U_{mn}(2N_1)$ keep the cluster state unchanged, so the probability amplitude for the target spin to return to its initial state is $\cos(2N_2\delta_2)$; (iii) if the cluster is initially in the other $d-4$ states, the DD control would keep the cluster state unchanged. So the sensor spin coherence for uncorrelated transitions is (See Appendix D)

$$L_{\text{Uncorre}}^{\text{dip}}(N_1, N_2) \approx \frac{1}{d}\left[d - 4 + 2\cos(2N_1\delta_1) + 2\cos(2N_2\delta_2)\right], \quad (6)$$

where $\delta_1 = |\beta_{mn}/\omega_{mn}|$ and $\delta_2 = |\beta_{pq}/\omega_{pq}|$. The case for two correlated transitions (with $|n\rangle = |p\rangle$) can be similarly discussed and the sensor coherence dip is (See Appendix D)

$$L_{Corre}^{dip}(N_1, N_2) \approx \frac{1}{d}\left[d - 3 + \cos(2N_1\delta_1) + \cos(2N_2\delta_2) + \cos(2N_1\delta_1)\cos(2N_2\delta_2)\right]. \quad (7)$$

The sensor coherence minima are different for uncorrelated and correlated transitions, being

$$\min\left(L_{Uncorre}^{dip}(N_1, N_2)\right) = (d-8)/d, \quad d \geq 4, \quad (8)$$

$$\min\left(L_{Corre}^{dip}(N_1, N_2)\right) = (d-4)/d, \quad d \geq 3. \quad (9)$$

Figure 3 shows the 2D sensor coherence as a function of $N_1, N_2$ for three different cases, i.e. type-V transition ($d = 3$), correlated and uncorrelated transitions in ladder type transitions ($d = 4$). For both correlated and uncorrelated transitions the sensor coherence dip oscillates periodically in the 2D space $(N_1, N_2)$ with the unit cell $N_{1c} \times N_{2c}$ ($N_{ic} = \pi/\delta_i$). The 2D sensor coherence presents different patterns for the correlated and uncorrelated transitions in the same target spin cluster. For correlated transitions in different target spin clusters, the sensor coherence shows similar patterns but with different coherence minima.

Higher-dimensional DD sequences can be similarly constructed for correlation sensing [24]. But it should be pointed out that the correlations between different transitions can already be fully determined by repeatedly applying the 2D DD sensing to different pairs of transitions.

## V. Discussion

To perform the 2D quantum sensing, we can take a shallow NV center in diamond near the surface as the sensor and choose the basis states $\{|+1\rangle, |-1\rangle\}$ of the NV center as the sensor states. The coherence of the double transition ($|+1\rangle \leftrightarrow |-1\rangle$) of the NV center can be generated, controlled and read out in a way similar to the coherence of the single transitions ($|0\rangle \leftrightarrow |\pm1\rangle$) by using composite microwave pulses [25, 26]. To determine whether two transitions in a single molecule are correlated or not, we can

measure the sensor coherence minimum with the 2D DD sequence $(N_1 = N_{1c}/2, N_2 = N_{2c}/2)$, which depends on whether the two transitions are correlated or uncorrelated [see Eq. (8) and Eq. (9)]. The pulse number period $N_{1c}$, $N_{2c}$ can be pre-deterimined by performing the 1D scan of the CPMG number for the two transitions, respectively [20]. From Eq. (6) and Eq. (7), we know that the minimum evolution time to observe the sensor coherence minima for the 2D sequence is $\pi^2/(2\delta_1\omega_1) + \pi^2/(2\delta_2\omega_2)$.

The measurement time for performing the 1D DD sensing scheme based on the NV center has been estimated in our previous work [20], now we estimate the measurement time for the 2D sensing scheme based on the NV center. The signal-to-noise ratio (SNR) for $K$ measurements of the NV spin coherence, with the spin projection noise and photon shot noise considered, is $\xi = F\sqrt{K}$ [27], where $F = \left[1 + 2(\alpha_0 + \alpha_1)/(\alpha_0 - \alpha_1)^2\right]^{-1/2}$ is the readout fidelity with $\alpha_0$, $\alpha_1$ being the mean number of detected photons per shot from the $|0\rangle$ and $|\pm1\rangle$ states of the NV center, respectively. To achieve a desired SNR $\xi$, the measurement cycle needs to be repeated for $K = \xi^2/F^2$ times. The total measurement time for observing the sensor coherence dip minimum caused by a single molecule is $T_k = K(t_k^D + t_{IR})$, where $t_k^D = \pi^2/(2\delta_1\omega_1) + \pi^2/(2\delta_2\omega_2)$ is the evolution time under the 2D DD control with $N_1 = N_{1c}/2, N_2 = N_{2c}/2$ and $t_{IR} \sim 1$ μs is the time for the initialization and readout of the NV center. For $\delta_1\omega_1, \delta_2\omega_2 \sim 2\pi \times 5$ kHz, $t_k^D = \pi^2/(2\delta_1\omega_1) + \pi^2/(2\delta_2\omega_2) \approx 0.31$ ms. The NV readout fidelity is $F \approx 0.03$ for typical fluorescence collection efficiencies. To reach an SNR $\xi = 10$, the measurement cycle needs to repeated for $K \approx 1.1 \times 10^5$ times and the total measurement time is $T_k \approx 34$ s for one data point ($N_1 = N_{1c}/2, N_2 = N_{2c}/2$) and $T_k \approx 9.2 \times 10^4$ s for all the the data in a unit cell $N_{1c} \times N_{2c}$ ($\Delta N_1, \Delta N_2 = 2$) in any of the three panels in Fig. 3(b). Recently the NV readout fidelity has been improved to about $F \approx 0.3$ by storing the NV electron spin state in an ancillary $^{15}$N nuclear spin [28]. With such a high fidelity, only $1.1 \times 10^3$ measurements are needed and the total measurement time would be reduced to about 0.34 s for one data point ($N_1 = N_{1c}/2, N_2 = N_{2c}/2$) and $T_k \approx 920$ s for all the the data in a unit cell $N_{1c} \times N_{2c}$ ($\Delta N_1, \Delta N_2 = 2$) in any of the three panels in Fig. 3(b).

## VI. Conclusion

We have proposed the concept of 2D DD-based quantum sensing and presented a realistic 2D DD sequence for correlation spectroscopy of single molecules. The 2D correlation spectra reveal correlations between different nuclear transitions, in a way similar to the 2D NMR correlation spectroscopy. Our work paves the way for structure and conformation analysis of single molecules labeled by nuclear spins or electron spins.

## Acknowledgements

This work was supported by Hong Kong Research Grants Council - Collaborative Research Fund CUHK4/CRF/12G and the Chinese University of Hong Kong Vice Chancellor's One-off Discretionary Fund.

## Appendix A: Derivation of Eq. (4) in the main text

The Hamiltonian representing a spin-1/2 sensor spin interacting with a single target spin cluster (or a single "molecule") is

$$H = S_z \beta + H_0, \tag{A1}$$

It can be recast into the eigenstates of the sensor states $|\pm\rangle$,

$$H = |+\rangle H^{(+)} \langle+| + |-\rangle H^{(-)} \langle-|, \tag{A2}$$

with the Hamiltonian of the target spin cluster conditioned on the sensor state

$$H^{(\pm)} = \pm \frac{1}{2}\beta + H_0, \tag{A3}$$

where $H_0 = \sum_{m=1}^{d} \varepsilon_m |m\rangle\langle m|$ is the free Hamiltonian for the target spin cluster with with $d$ denoting the number of eigenstates $\{|m\rangle\}$, and $\beta = \frac{1}{2}\sum_{m,n}(\beta_{mn}|m\rangle\langle n| + \text{H.c.})$ is the noise operator from the target spin cluster with $\beta_{mn} = \langle m|\beta|n\rangle$. In the interaction picture set by $H_0$, the time-dependent noise operator is

$$\beta(t) = e^{iH_0 t}\beta e^{-iH_0 t} = \frac{1}{2}\sum_{m,n}(\beta_{mn} e^{i\omega_{mn} t}|m\rangle\langle n| + \text{H.c.}), \tag{A4}$$

where $\omega_{mn} = \varepsilon_m - \varepsilon_n$ is the frequency for the transition $|m\rangle \leftrightarrow |n\rangle$. The time evolution of the target spin cluster conditioned on the sensor state is

$$U_N^{(\pm)}(t) = e^{-iH_0 t} T e^{\mp \frac{i}{2}\int_0^t f(t')\beta(t')dt'}, \tag{A5}$$

where $f(t) = (-1)^p$ for $[t_p, t_{p+1}]$ [$t_p = (2p-1)t/(2N)$ for $p = 1, \cdots, N$ and $t_0 = 0, t_{N+1} = t$] is the DD modulation function and $T$ is the time-ordering operator.

Now we use the Magnus expansion [20, 29] to get a simple formula for $U_N^{(\pm)}(t)$. which is valid for weak sensor-target coupling. According to the Magnus expansion, a general time-dependent evolution operator can expanded as

$$U(t) = T e^{-i\int_0^t H(t')dt'} = \exp\left(\sum_{l=1}^{\infty} \Omega_l(t)\right), \tag{A6}$$

with the first-order and second-order Magnus terms

$$\Omega_1(t) = -i\int_0^t H(t')dt', \tag{A7}$$

$$\Omega_2(t) = -\frac{1}{2}\int_0^t dt_1 \int_0^{t_1} dt_2 [H(t_1), H(t_2)]. \tag{A8}$$

For the time evolution operator $T e^{\mp \frac{i}{2}\int_0^t f(t')\beta(t')dt'}$ in Eq. (A5), the first-order Magnus term is

$$\Omega_1(t) = -\frac{i}{2} \sum_{m>n} \left( \beta_{mn} \frac{F_N(\omega_{mn}, t) e^{i\xi_{mn}t}}{\omega_{mn}} |m\rangle\langle n| + \text{H.c.} \right), \tag{A9}$$

where we have defined the filter function $F_N(\omega_{mn}, t)$ for $N$-pulse DD as

$$F_N(\omega_{mn}, t) = \omega_{mn} \left| \int_0^t f(t') e^{i\omega_{mn}t'} dt' \right|, \tag{A10}$$

and the phase $\xi_{mn}$ as

$$e^{i\xi_{mn}t} = \left( \omega_{mn} \int_0^t f(t') e^{i\omega_{mn}t'} dt' \right) \Big/ F_N(\omega_{mn}, t). \tag{A11}$$

Here the diagonal terms with $m = n$ in $\beta(t)$ are averaged out by the DD control, so there is no diagonal terms in $\Omega_1(t)$. In the weak-coupling regime, i.e. $|\beta_{mn}| \ll |\omega_{mn}|$, it has been demonstrated that $\|\Omega_2(t)\| \ll \|\Omega_1(t)\|$ in the short-time scale [20, 29].

For the CPMG-$N_1$ control, the filter function is [20, 21]

$$F_{N_1}(\omega, t) = \begin{cases} 4\sin^2\left(\frac{\omega t}{4N_1}\right) \left| \cos\left(\frac{\omega t}{2}\right) \cos^{-1}\left(\frac{\omega t}{2N_1}\right) \right|, & \text{for odd } N_1, \\ 4\sin^2\left(\frac{\omega t}{4N_1}\right) \left| \sin\left(\frac{\omega t}{2}\right) \cos^{-1}\left(\frac{\omega t}{2N_1}\right) \right|, & \text{for even } N_1. \end{cases} \tag{A12}$$

Under the resonance condition for $\omega_{mn}$, i.e., when $t = 2N_1\tau_1 = 2N_1\tau_{mn} = \pi(2c-1)N_1/\omega_{mn}$ ($c = 1, 2, \cdots$), we have $F(\omega_{mn}, 2N_1\tau_{mn}) = 2N_1$ and $\xi_{mn} = 0$. Here we have assumed that different target spin transitions have no joint contributions to the sensor coherence dip, that is, $F_{N_1}(\omega_{pq}, 2N_1\tau_{mn}) \ll F_{N_1}(\omega_{mn}, 2N_1\tau_{mn})$ for $p, q \neq m, n$. Then the first-order Magnus term can be simplified as

$$\Omega_1(2N_1\tau_{mn}) = -iN_1 \left( \beta_{mn} |m\rangle\langle n| + \text{H.c.} \right)/\omega_{mn}. \tag{A13}$$

With the definition

$$U_{mn}(N_1) = \exp(\Omega_1) = \exp\left[-iN_1\left(\beta_{mn}|m\rangle\langle n| + \text{H.c.}\right)/\omega_{mn}\right], \tag{A14}$$

the time evolution operator $U_{N_1}^{(\pm)}(t)$ is approximated as

$$U_{N_1}^{(\pm)}(2N_1\tau_{mn}) \approx \exp(-2iH_0 N_1\tau_{mn}) U_{mn}(\pm N_1), \tag{A15}$$

and the sensor spin coherence dip is

$$L^{\text{dip}}(N_1) \approx \frac{1}{d}\text{Tr}\left[\left(U_{N_1}^{(-)}(2N_1\tau_{mn})\right)^\dagger U_{N_1}^{(+)}(2N_1\tau_{mn})\right] = \frac{1}{d}\text{Tr}\left[U_{mn}(2N_1)\right]. \tag{A16}$$

Since the operator $2\Omega_1$ in $U_{mn}(2N_1)$ has only two non-zero eigenvalues: $\pm 2iN_1\delta_1 = \pm 2iN_1|\beta_{mn}/\omega_{mn}|$, the sensor coherence dip is

$$L^{\text{dip}}(N_1) \approx \frac{1}{d}\text{Tr}\left[U_{mn}(2N_1)\right] = \frac{1}{d}\left[d - 2 + 2\cos(2N_1\delta_1)\right], \tag{A17}$$

which is Eq. (4) in the main text.

## Appendix B: Sensing uncorrelated transitions from independent "molecules"

The Hamiltonian representing a spin-1/2 sensor spin weakly coupled to $M$ target clusters of nuclear spins (each cluster representing a "molecule") is

$$H = S_z \sum_{k=1}^{M} \beta^{(k)} + \sum_{k=1}^{M} H_0^{(k)}, \tag{B1}$$

where $H_0^{(k)} = \sum_{m=1}^{d_k} \varepsilon_m^{(k)} |m\rangle_k \langle m|$ is the free Hamiltonian for the $k$-th target cluster with $d_k$ denoting the number of eigenstates, and $\beta^{(k)} = \frac{1}{2}\sum_{m,n}\left(\beta_{mn}^{(k)}|m\rangle_k\langle n| + \text{H.c.}\right)$ is the noise operator from the $k$-th nuclear spin cluster which induces the nuclear spin transition $|m\rangle_k \leftrightarrow |n\rangle_k$. The time-dependent noise operator is

$$\beta^{(k)}(t)=e^{iH_0^{(k)}t}\beta^{(k)}e^{-iH_0^{(k)}t} = \frac{1}{2}\sum_{m,n}\left(\beta_{mn}^{(k)}e^{i\omega_{mn}^{(k)}t}|m\rangle_{kk}\langle n|+\text{H.c.}\right). \tag{B2}$$

The time evolution of the target spin clusters conditioned on the sensor state is

$$U_N^{(\pm)}(t) = \prod_{k=1}^{M}U_{k,N}^{(\pm)}(t) = \prod_{k=1}^{M}e^{-iH_0^{(k)}t}Te^{\mp\frac{i}{2}\int_0^t f(t')\beta^{(k)}(t')dt'}. \tag{B3}$$

The first-order Magnus term of $Te^{\mp\frac{i}{2}\int_0^t f(t')\beta^{(k)}(t')dt'}$ is

$$\Omega_1^{(k)}(t)=-\frac{i}{2}\sum_{m>n}\left(\beta_{mn}^{(k)}\frac{F\left(\omega_{mn}^{(k)},t\right)}{\omega_{mn}^{(k)}}e^{i\xi_{mn}^{(k)}t}|m\rangle_{kk}\langle n|+\text{H.c.}\right), \tag{B4}$$

where we have defined the filter function $F_N(\omega_{mn},t)$ for the CPMG-$N$ DD as

$$F_N(\omega_{mn}^{(k)},t) = \omega_{mn}^{(k)}\left|\int_0^t f(t')e^{i\omega_{mn}^{(k)}t'}dt'\right|, \tag{B5}$$

and for CPMG control the phase $\xi_{mn}^{(k)}(N,2\tau)$ is a joint function of the pulse number $N$ and pulse delay $2\tau$ as

$$e^{i\xi_{mn}^{(k)}(N,2\tau)}=\left(\omega_{mn}^{(k)}\int_0^t f(t')e^{i\omega_{mn}^{(k)}t'}dt'\right)\Big/F_N\left(\omega_{mn}^{(k)},t\right). \tag{B6}$$

The filter function for the 2D DD sequence ($t = 2N_1\tau_1 + 2N_2\tau_2$, $N = N_1 + N_2$) is

$$F_{(N_1,N_2)}(\omega_{mn}^{(k)},t) = \left|F_{N_1}(\omega_{mn}^{(k)},2N_1\tau_1)e^{i\xi_{mn}^{(k)}(N_1,2\tau_1)} + F_{N_2}(\omega_{mn}^{(k)},2N_2\tau_2)e^{i\xi_{mn}^{(k)}(N_2,2\tau_2)}\right|. \tag{B7}$$

We choose a transition $|m\rangle_1 \leftrightarrow |n\rangle_1$ from target 1 and another transition $|p\rangle_2 \leftrightarrow |q\rangle_2$ from target 2. Under the double-resonance condition ($2\tau_1=\pi/\omega_{mn}^{(1)}$, $2\tau_2=\pi/\omega_{pq}^{(2)}$),

$$F_{N_1}(\omega_{mn}^{(1)}, 2N_1\tau_1) = 2N_1 \gg F_{N_2}(\omega_{mn}^{(1)}, 2N_2\tau_2), \tag{B8}$$

$$F_{N_2}(\omega_{pq}^{(2)}, 2N_2\tau_2) = 2N_2 \gg F_{N_1}(\omega_{pq}^{(2)}, 2N_1\tau_1), \tag{B9}$$

so the filter function for the 2D DD sequence can be approximated as

$$F_{(N_1,N_2)}(\omega_{mn}^{(1)}, 2N_1\tau_1 + 2N_2\tau_2) \approx 2N_1, \tag{B10}$$

$$F_{(N_1,N_2)}(\omega_{pq}^{(2)}, 2N_1\tau_1 + 2N_2\tau_2) \approx 2N_2. \tag{B11}$$

Then the first-order Magnus terms become

$$\Omega_1^{(1)}(2N_1\tau_{mn} + 2N_2\tau_{pq}) = -iN_1\left(\beta_{mn}^{(1)}|m\rangle_{11}\langle n| + \text{H.c.}\right)/\omega_{mn}^{(1)}, \tag{B12}$$

$$\Omega_1^{(2)}(2N_1\tau_{mn} + 2N_2\tau_{pq}) = -iN_2\left(\beta_{pq}^{(2)}|p\rangle_{22}\langle q| + \text{H.c.}\right)/\omega_{pq}^{(2)}. \tag{B13}$$

With the definition

$$U_{1,mn}(N_1) = \exp(\Omega_1^{(1)}) = \exp\left[-iN_1\left(\beta_{mn}^{(1)}|m\rangle_{11}\langle n| + \text{H.c.}\right)/\omega_{mn}^{(1)}\right], \tag{B14}$$

$$U_{2,pq}(N_2) = \exp(\Omega_1^{(2)}) = \exp\left[-iN_2\left(\beta_{pq}^{(2)}|p\rangle_{22}\langle q| + \text{H.c.}\right)/\omega_{pq}^{(2)}\right], \tag{B15}$$

the sensor coherence dip is

$$\begin{aligned}L_{1,2}^{\text{dip}}(N_1, N_2) &\approx \frac{1}{d_1 d_2}\text{Tr}_1\left[U_{1,mn}(2N_1)\right]\text{Tr}_2\left[U_{2,pq}(2N_2)\right] \\ &= \frac{1}{d_1 d_2}\left[d_1 - 2 + 2\cos(2N_1\delta_1)\right]\left[d_2 - 2 + 2\cos(2N_2\delta_2)\right].\end{aligned} \tag{B16}$$

## Appendix C: Derivation of Eq. (5) in the main text

For a 2D DD sequences resonant with two different transitions in a single target spin cluster ($2\tau_1 = 2\tau_{mn} = \pi/\omega_{mn}$, $2\tau_2 = 2\tau_{pq} = \pi/\omega_{pq}$), the time evolution operator of the target spin is the product of the two time evolution operators for CMPG-$N_1$ with $\tau_{mn}$ and CPMG-$N_2$ with $\tau_{pq}$ respectively,

$$U^{(\pm)}_{(N_1,N_2)}\left(2N_1\tau_{mn}+2N_2\tau_{pq}\right)$$
$$=e^{-2iH_0N_2\tau_{pq}}U^{(\pm)}_{N_2}\left(2N_2\tau_{pq}\right)e^{-2iH_0N_1\tau_{mn}}U^{(\pm)}_{N_1}\left(2N_1\tau_{mn}\right) \quad (C1)$$
$$=e^{-2iH_0N_2\tau_{pq}}U_{pq}\left(\pm N_2\right)e^{-2iH_0N_1\tau_{mn}}U_{mn}\left(\pm N_1\right).$$

Here we have assumed that $N_1$ is even (when $N_1$ is odd, we should replace $U_{pq}(\pm N_2)$ with $U_{pq}(\mp N_2)$ in Eq. (C1) or keep using this formula but modify the phase of $\beta_{pq}$ correspondingly). The sensor coherence dip is

$$L^{\mathrm{dip}}(N_1,N_2)=\frac{1}{d}\mathrm{Tr}\left[\left(U^{(-)}_{(N_1,N_2)}\right)^\dagger U^{(+)}_{(N_1,N_2)}\right]$$
$$\approx \frac{1}{d}\mathrm{Tr}\left[U_{mn}(N_1)e^{2iH_0N_1\tau_{mn}}U_{pq}(2N_2)e^{-2iH_0N_1\tau_{mn}}U_{mn}(N_1)\right] \quad (C2)$$
$$=\frac{1}{d}\mathrm{Tr}\left[U_{pq}(2N_2)e^{-2iH_0N_1\tau_{mn}}U_{mn}(2N_1)e^{2iH_0N_1\tau_{mn}}\right],$$

with the term

$$e^{-2iH_0N_1\tau_{mn}}U_{mn}(2N_1)e^{2iH_0N_1\tau_{mn}}=\sum_{a,b}e^{-2i\omega_{ab}N_1\tau_{mn}}|a\rangle\langle a|U_{mn}(2N_1)|b\rangle\langle b|. \quad (C3)$$

The diagonal element $\langle a|e^{-2iH_0N_1\tau_{mn}}U_{mn}(2N_1)e^{2iH_0N_1\tau_{mn}}|a\rangle$ obviously equals to $\langle a|U_{mn}(2N_1)|a\rangle$, and the non-zero off-diagonal elements are

$$\langle m|e^{-2iH_0N_1\tau_{mn}}U_{mn}(2N_1)e^{2iH_0N_1\tau_{mn}}|n\rangle=(-1)^{N_1}\langle m|U_{mn}(2N_1)|n\rangle, \quad (C4)$$

$$\langle n|e^{-2iH_0N_1\tau_{mn}}U_{mn}(2N_1)e^{2iH_0N_1\tau_{mn}}|m\rangle=(-1)^{N_1}\langle n|U_{mn}(2N_1)|m\rangle, \quad (C5)$$

which only change the sign when the pulse number $N_1$ is odd. When $N_1$ is even, the sensor coherence dip is

$$L^{\mathrm{dip}}(N_1,N_2)\approx\frac{1}{d}\mathrm{Tr}\left[U_{pq}(2N_2)U_{mn}(2N_1)\right], \quad (C6)$$

which is just Eq. (5) in the main text. When $N_1$ is odd, the sensor coherence dip is

$$L^{\text{dip}}(N_1, N_2) \approx \frac{1}{d}\text{Tr}\left[U_{pq}(-2N_2)U_{mn}(-2N_1)\right],$$ which is equivalent to Eq. (C6).

## Appendix D: Derivation of Eq. (6) and (7) in the main text

First we consider a simple case with $d = 4$. Typically $\beta_{mn}$ is complex and we denote $\beta_{mn}/\omega_{mn} = \delta_{mn}e^{i\kappa_{mn}}$.

(i) If the two transitions are uncorrelated ($m=1, n=2, p=3, q=4$):

$$U_{12}(2N_1) = \begin{bmatrix} \cos(2N_1\delta_{12}) & -ie^{i\kappa_{12}}\sin(2N_1\delta_{12}) & 0 & 0 \\ -ie^{-i\kappa_{12}}\sin(2N_1\delta_{12}) & \cos(2N_1\delta_{12}) & 0 & 0 \\ 0 & 0 & 1 & 0 \\ 0 & 0 & 0 & 1 \end{bmatrix}. \quad \text{(D1)}$$

$$U_{34}(2N_2) = \begin{bmatrix} 1 & 0 & 0 & 0 \\ 0 & 1 & 0 & 0 \\ 0 & 0 & \cos(2N_2\delta_{34}) & -ie^{i\kappa_{34}}\sin(2N_2\delta_{34}) \\ 0 & 0 & -ie^{-i\kappa_{34}}\sin(2N_2\delta_{34}) & \cos(2N_2\delta_{34}) \end{bmatrix}. \quad \text{(D2)}$$

The sensor coherence dip is

$$L^{\text{dip}}_{\text{Uncorre}}(N_1, N_2) \approx \frac{1}{4}\text{Tr}\left[U_{12}(2N_1)U_{34}(2N_2)\right] = \frac{1}{2}\left[\cos(2N_1\delta_{12}) + \cos(2N_2\delta_{34})\right]. \quad \text{(D3)}$$

(ii) If the two transitions are correlated ($m=1, n=2, p=2, q=3$):

$$U_{12}(2N_1) = \begin{bmatrix} \cos(2N_1\delta_{12}) & -ie^{i\kappa_{12}}\sin(2N_1\delta_{12}) & 0 & 0 \\ -ie^{-i\kappa_{12}}\sin(2N_1\delta_{12}) & \cos(2N_1\delta_{12}) & 0 & 0 \\ 0 & 0 & 1 & 0 \\ 0 & 0 & 0 & 1 \end{bmatrix}, \quad \text{(D4)}$$

$$U_{23}(2N_2) = \begin{bmatrix} 1 & 0 & 0 & 0 \\ 0 & \cos(2N_2\delta_{23}) & -ie^{i\kappa_{23}}\sin(2N_2\delta_{23}) & 0 \\ 0 & -ie^{-i\kappa_{23}}\sin(2N_2\delta_{23}) & \cos(2N_2\delta_{23}) & 0 \\ 0 & 0 & 0 & 1 \end{bmatrix}. \tag{D5}$$

The sensor coherence dip is

$$\begin{aligned} L_{Corre}^{dip}(N_1, N_2) &\approx \frac{1}{4}\text{Tr}\left[U_{12}(2N_1)U_{23}(2N_2)\right] \\ &= \frac{1}{4}\left[1 + \cos(2N_1\delta_{12}) + \cos(2N_2\delta_{23}) + \cos(2N_1\delta_{12})\cos(2N_2\delta_{23})\right]. \end{aligned} \tag{D6}$$

The above derivation can be easily generalized to an arbitrary $d$ to get the sensor coherence dip for uncorrelated and correlated transitions,

$$L_{Uncorre}^{dip}(N_1, N_2) \approx \frac{1}{d}\left[d - 4 + 2\cos(2N_1\delta_1) + 2\cos(2N_2\delta_2)\right], \tag{D7}$$

$$L_{Corre}^{dip}(N_1, N_2) \approx \frac{1}{d}\left[d - 3 + \cos(2N_1\delta_1) + \cos(2N_2\delta_2) + \cos(2N_1\delta_1)\cos(2N_2\delta_2)\right], \tag{D8}$$

where $\delta_1 = \delta_{mn}$ and $\delta_2 = \delta_{pq}$. These two formula are Eq. (6) and (7) in the main text.

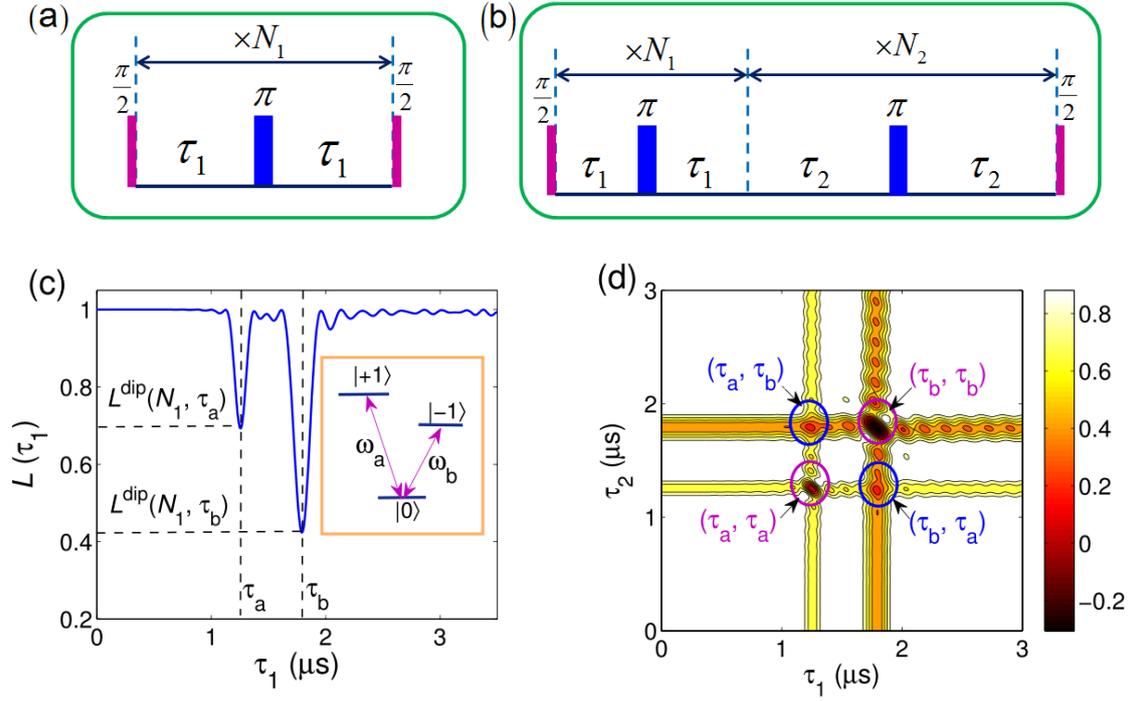

**Fig. 1**. (a) One-dimensional DD sequence. (b) Two-dimensional DD sequence. Here the $l$ - dimensional DD sequence contains $l$ CPMG sub-sequences with different pulse intervals $2\tau_i$ and pulse numbers $N_i$. The $l$-dimensional DD is resonant with $l$ different transitions when the pulse intervals are such that $2\tau_i = \pi/\omega_i$ for $i = 1, 2, \cdots, l$. (c) Sensor coherence as a function of the pulse interval $\tau_1$ under a one-dimensional DD control with $N_1 = 20$. The spin-1/2 sensor is weakly coupled to a single target spin-1 **J** via the coupling Hamiltonian $H = \lambda S_z J_x + (\omega_a + \omega_b) J_z^2 / 2 + (\omega_a - \omega_b) J_z / 2$ with $\lambda = 5\sqrt{2}$ kHz and the two transition frequencies $\omega_a = 0.20$ MHz and $\omega_b = 0.14$ MHz (see inset). The sensor coherence shows sharp dips when the pulse interval matches the nuclear spin transition frequency $\tau_{a/b} = \pi/(2\omega_{a/b})$. (d) Sensor coherence as a function of two pulse intervals $\tau_1, \tau_2$ under a 2D DD control with $N_1 = N_2 = 20$. The sensor coherence shows four dips in the $(\tau_1, \tau_2)$ space. The two diagonal dips at $(\tau_a, \tau_a), (\tau_b, \tau_b)$ correspond to the one-dimensional resonant DD case as that in (c), while the other two symmetrical off-diagonal dips at $(\tau_a, \tau_b), (\tau_b, \tau_a)$ contain information about the correlation between the two different transitions. Here the sensor is coupled to the same target spin as that in (c).

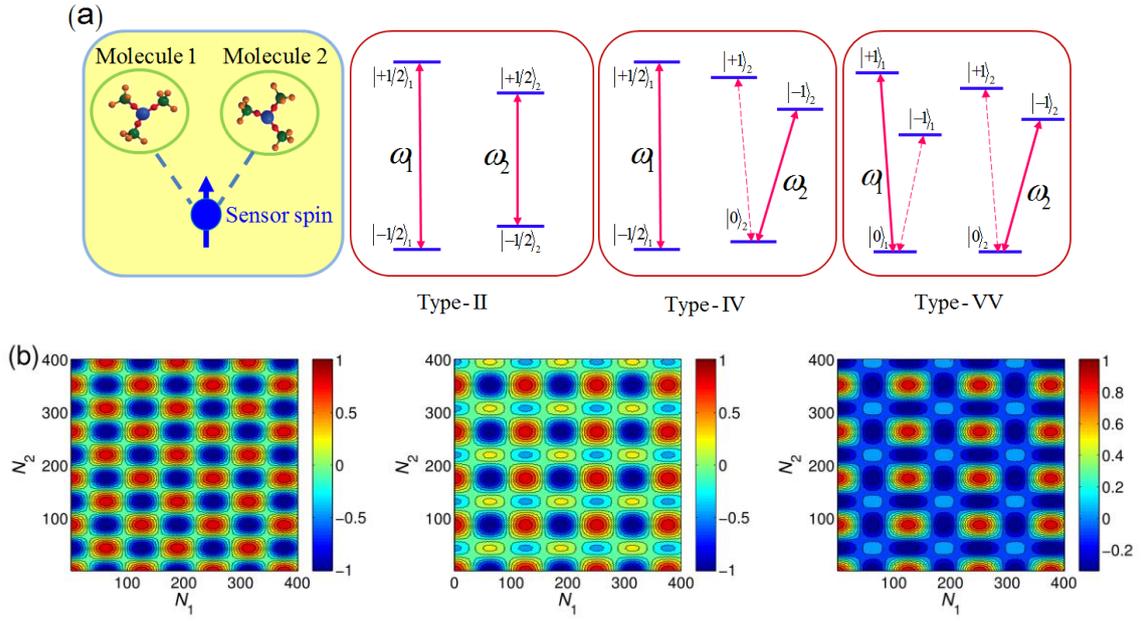

**Fig. 2**. Two-dimensional quantum sensing of two independent clusters of nuclear spins. (a) Schematic illustration of the sensor coupled to two target "molecules" and different types of transitions from the two targets. Type-II transitions represent two independent target spin-1/2's, type-IV transitions represent a target spin-1/2 and a target spin-1, and type-VV transitions represent two independent target spin-1's. The solid (dashed) arrows are nuclear spin transitions resonant (off-resonant) with the DD. (b) Sensor coherence dip as a function of two CPMG pulse numbers $N_1$ and $N_2$ for type-II, type-IV and type-VV transitions in (a) correspondingly. The parameters are such that $\omega_1 = 0.20$ MHz, $\omega_2 = 0.14$ MHz, $\delta_1 = 0.025$, and $\delta_2 = 0.036$.

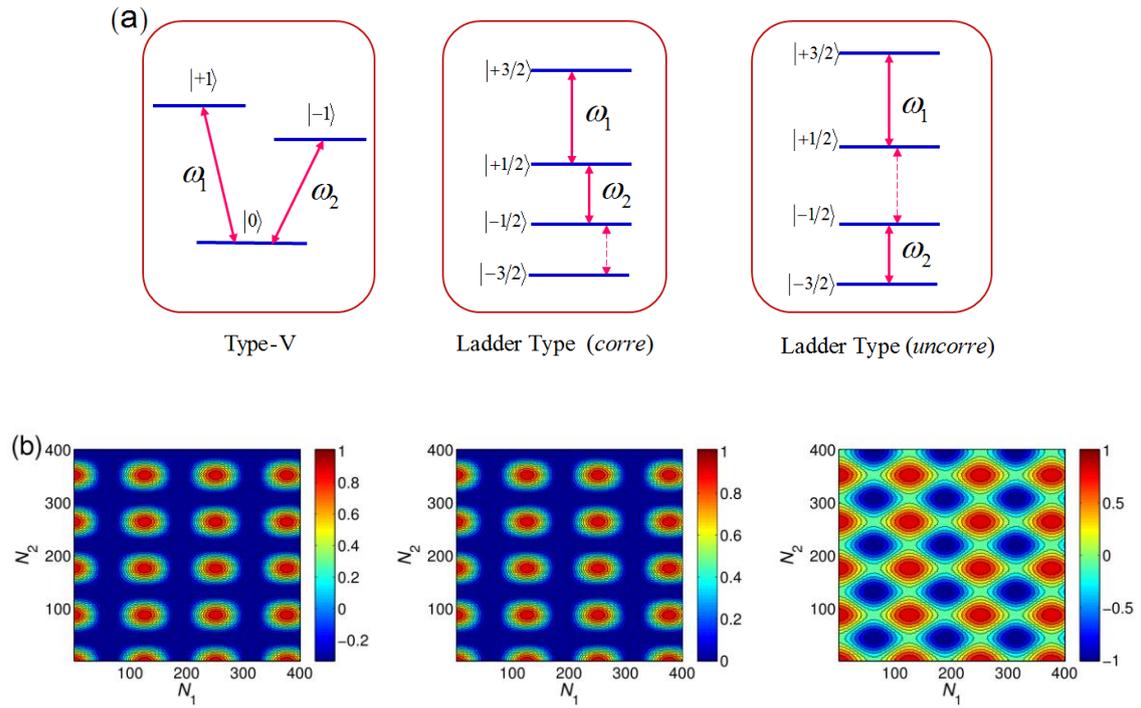

**Fig. 3**. Two-dimensional quantum sensing of correlations in a single nuclear spin cluster. (a) Schematic illustration of different correlation types of two transitions in a single nuclear spin cluster. Type-V transitions represent two correlated transitions in a spin-1 cluster, ladder type (*corre*) transitions represent two correlated transitions in a spin-3/2 cluster, and ladder type (*uncorre*) transitions represent two uncorrelated transitions in a spin-3/2 cluster. The solid (dashed) arrows are nuclear spin transitions resonant (off-resonant) with the DD. (b) Sensor coherence dip as a function of two CPMG pulse numbers $N_1$ and $N_2$ for type-V, ladder type (*corre*) and ladder type (*uncorre*) transitions in (a) correspondingly. The parameters are such that $\omega_1 = 0.20$ MHz, $\omega_2 = 0.14$ MHz, $\delta_1 = 0.025$, and $\delta_2 = 0.036$.

**Supplementary Information for "Proposal for Quantum Sensing Based on Two-dimensional Dynamical Decoupling: NMR Correlation Spectroscopy of Single Molecules"**

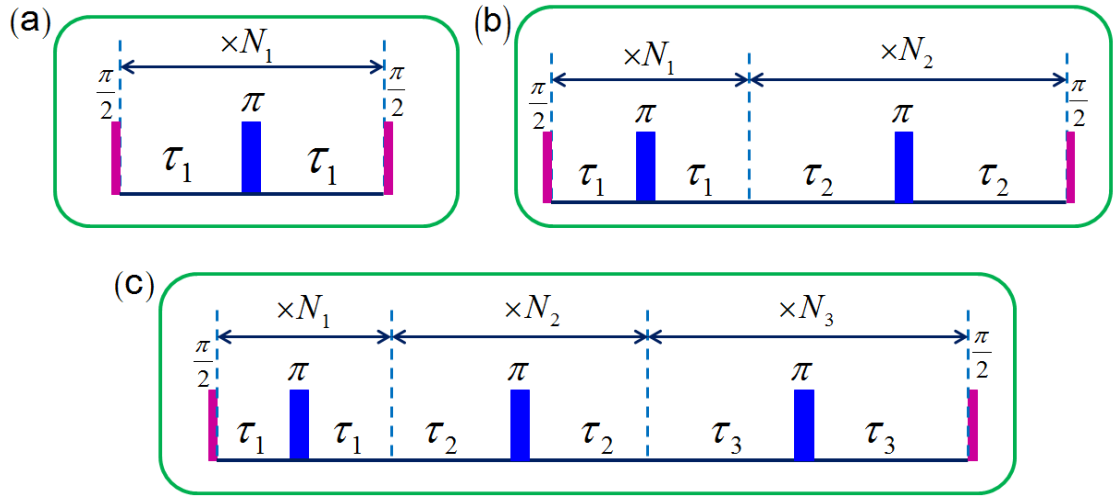

**Fig. S1**. Multi-dimensional DD sequences. (a) One-dimensional DD sequence. (b) Two-dimensional DD sequence. (c) Three-dimensional DD sequence. Here the $l$-dimensional DD sequence contains $l$ CPMG sub-sequences with different pulse intervals $2\tau_i$ and pulse numbers $N_i$. The $l$-dimensional DD is resonant with $l$ different transitions when the pulse intervals are such that $2\tau_i = \pi/\omega_i$ for $i = 1, 2, \cdots, l$.

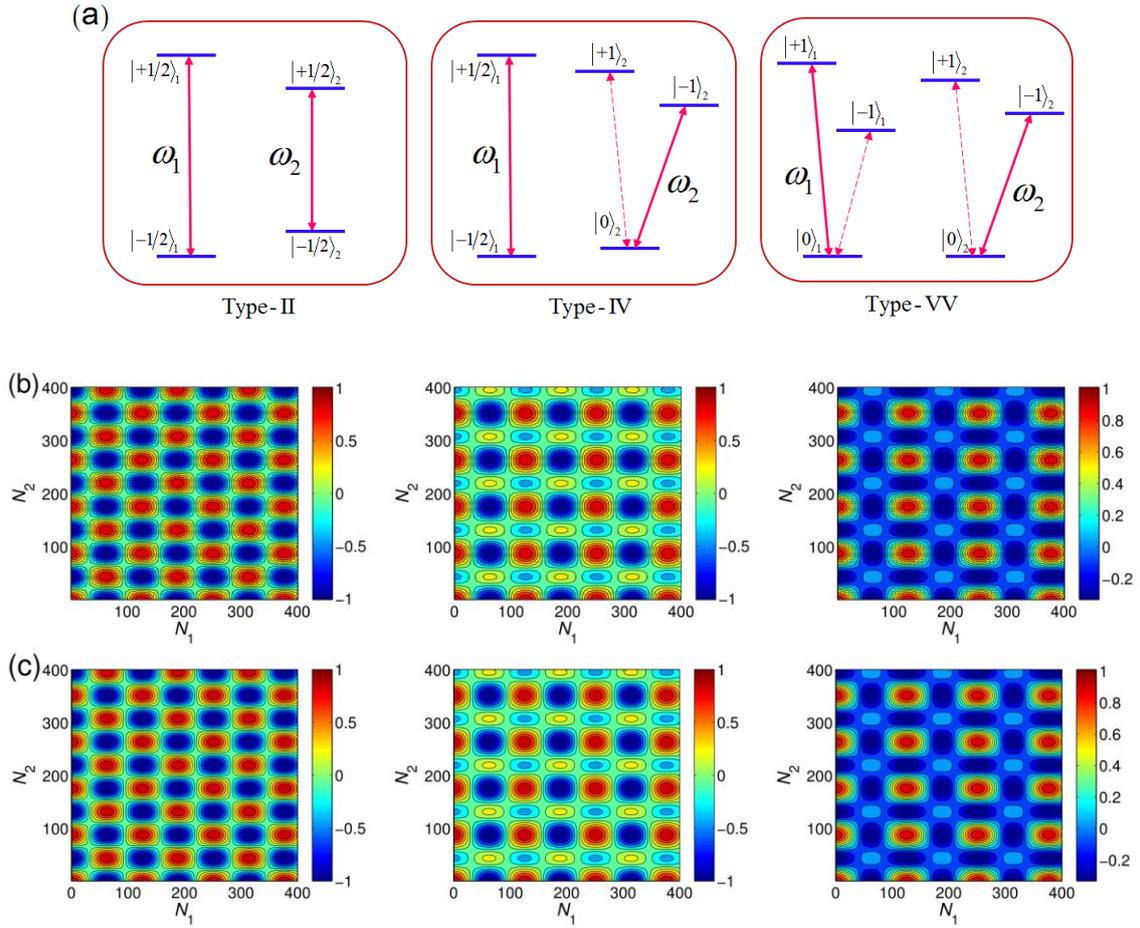

**Fig. S2**. Comparison between exact and analytical results for 2D quantum sensing of uncorrelated transitions from two independent nuclear spins. (a) Schematic illustration of the sensor coupled to two target "molecules" and different types of transitions from the two targets. Type-II transitions represent two independent target spin-1/2's, type-IV transitions represent a target spin-1/2 and a target spin-1, and type-VV transitions represent two independent target spin-1's. The solid (dashed) arrows are nuclear spin transitions resonant (off-resonant) with the DD. (b) Exactly calculated sensor coherence dip as a function of two CPMG pulse numbers $N_1$ and $N_2$ for type-II, type-IV and type-VV transitions in (a) correspondingly. (c) is similar to (b) but from the analytical formula in Eq. (B16) in the main text. The parameters are such that $\omega_1 = 0.20$ MHz, $\omega_2 = 0.14$ MHz, $\delta_1 = 0.025$, $\delta_2 = 0.036$ (the same as those in Fig. 2 in the main text).

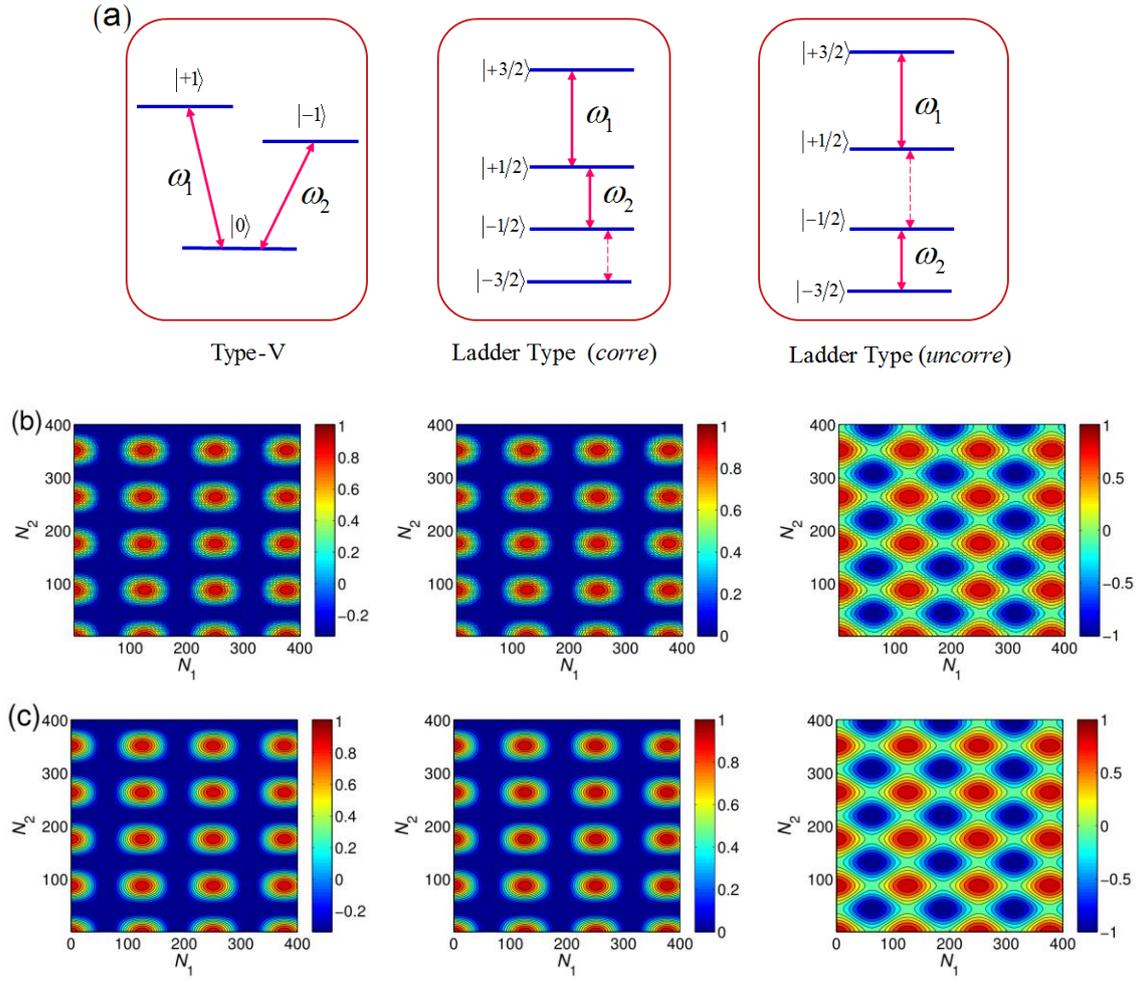

**Fig. S3**. Comparison between exact and analytical results for 2D quantum sensing of correlated and uncorrelated transitions in a nuclear spin cluster. (a) Schematic illustration of different correlation types of two transitions in a single nuclear spin cluster. Type-II transitions represent two correlated transitions in a spin-1 cluster, ladder type (*corre*) transitions represent two correlated transitions in a spin-3/2 cluster, and ladder type (*uncorre*) transitions represent two uncorrelated transitions in a spin-3/2 cluster. The solid (dashed) arrows are nuclear spin transitions resonant off-resonant) with the DD. (b) Exactly calculated sensor coherence dip as a function of two CPMG pulse numbers $N_1$ and $N_2$ for type-V, ladder type (*corre*) and ladder type (*uncorre*) transitions in (a) correspondingly. (c) is similar to (b) but from the analytical formula in Eq. (6) and Eq. (7) in the main text. The parameters are such that $\omega_1 = 0.20$ MHz, $\omega_2 = 0.14$ MHz, $\delta_1 = 0.025$, $\delta_2 = 0.036$ (the same as those in Fig. 3 in the main text).

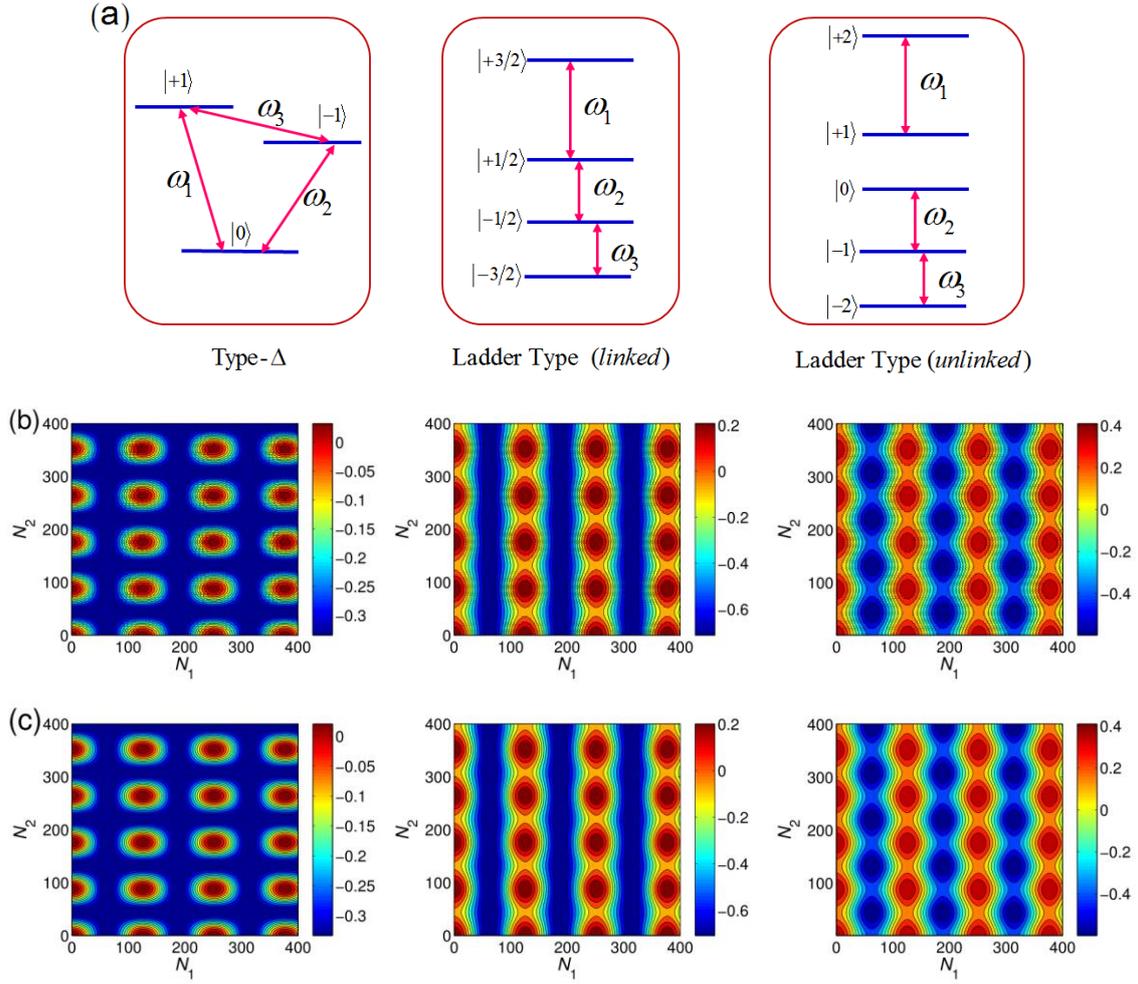

**Fig. S4**. Comparison between exact and analytical results for three-dimensional quantum sensing of different types of transitions in a nuclear spin cluster. (a) Schematic illustration of different correlation types of three transitions from a single nuclear spin cluster. Type-$\Delta$ transitions represent three correlated transitions in a spin-1 cluster, ladder type (*linked*) transitions represent three correlated transitions in a spin-3/2, and ladder type (*unlinked*) transitions represent one uncorrelated transition and two correlated transitions in a spin-3/2 cluster. (b) Exactly calculated sensor coherence dip as a function of two CPMG pulse numbers $N_1, N_2$ with $N_3 = 12$ for type-$\Delta$, ladder type (*linked*) and ladder type (*unlinked*) transitions in (a) correspondingly. (c) is similar to (b) but from the analytical formula in Eq. (S6)-(S8). The parameters are such that $\omega_1 = 0.20$ MHz, $\omega_2 = 0.14$ MHz, $\omega_3 = 0.06$ MHz, $\delta_1 = 0.025$, $\delta_2 = 0.036$, $\delta_3 = 0.083$.

# I. Three-dimensional DD-based quantum sensing

Higher dimensional sensing can be introduced to further differentiate different types of correlations in the nuclear spin clusters. We demonstrate this idea with the three-dimensional (3D) DD-based quantum sensing. The 3D sequence contains three subsequent sets of CPMG sequences with pulse intervals $2\tau_1, 2\tau_2, 2\tau_3$ and pulse numbers $N_1, N_2, N_3$ [Fig. S1(c)]. The sensor spin is flipped at times $t_{p_1} = (2p_1 - 1)\tau_1$, $t_{N_1+p_2} = 2N_1\tau_1 + (2p_2 - 1)\tau_2$, $t_{N_1+N_2+p_3} = 2N_1\tau_1 + 2N_2\tau_2 + (2p_3 - 1)\tau_3$ with $p_i = 1, 2, \cdots, N_i$ ($i = 1, 2, 3$). When the three pulse intervals match the frequencies of three different target spin transitions ($2\tau_i = \pi/\omega_i$), the sensor coherence as a function of $N_1$, $N_2$ and $N_3$ directly reveals the multiple correlations between the transitions.

For example, we consider the case that the sensor spin is coupled with three independent target "molecules" ($M = 3$) and the DD is resonant with three transitions $|m\rangle_1 \leftrightarrow |n\rangle_1$, $|p\rangle_2 \leftrightarrow |q\rangle_2$, $|r\rangle_3 \leftrightarrow |s\rangle_3$ from target 1,2,3 (with transition frequencies $\omega_{mn}^{(1)}, \omega_{pq}^{(2)}, \omega_{rs}^{(3)}$ and transition matrix elements $\beta_{mn}^{(1)}, \beta_{pq}^{(2)}, \beta_{rs}^{(3)}$ in turn). Under the resonance condition, the sensor spin coherence as a function of the pulse numbers $N_1$, $N_2$, and $N_3$ is

$$L_{1,2,3}^{\text{dip}}(N_1, N_2, N_3) \approx \frac{1}{d_1 d_2 d_3}\left[d_1 - 2 + 2\cos(2N_1\delta_1)\right]\left[d_2 - 2 + 2\cos(2N_2\delta_2)\right] \times \left[d_3 - 2 + 2\cos(2N_3\delta_3)\right], \quad (S1)$$

where $\delta_1 = \left|\beta_{mn}^{(1)}/\omega_{mn}^{(1)}\right|$, $\delta_2 = \left|\beta_{pq}^{(2)}/\omega_{pq}^{(2)}\right|$, and $\delta_3 = \left|\beta_{rs}^{(3)}/\omega_{rs}^{(3)}\right|$.

For the 3D DD sequences resonant with three different transitions $|m\rangle \leftrightarrow |n\rangle$, $|p\rangle \leftrightarrow |q\rangle$ and $|r\rangle \leftrightarrow |s\rangle$ in the same target "molecule" ($2\tau_1 = 2\tau_{mn} = \pi/\omega_{mn}, 2\tau_2 = 2\tau_{pq} = \pi/\omega_{pq}, 2\tau_3 = 2\tau_{rs} = \pi/\omega_{rs}$), the time evolution operator of the target spin cluster is the product of the three time evolution operators for CMPG-$N_1$ with $\tau_{mn}$, CPMG-$N_2$ with $\tau_{pq}$ and CPMG-$N_3$ with $\tau_{rs}$ correspondingly,

$$U^{(\pm)}_{(N_1,N_2,N_3)}\left(2N_1\tau_{mn}+2N_2\tau_{pq}+2N_3\tau_{rs}\right)$$
$$=e^{-2iH_0N_3\tau_{rs}}U^{(\pm)}_{N_3}\left(2N_3\tau_{rs}\right)e^{-2iH_0N_2\tau_{pq}}U^{(\pm)}_{N_2}\left(2N_2\tau_{pq}\right)e^{-2iH_0N_1\tau_{mn}}U^{(\pm)}_{N_1}\left(2N_1\tau_{mn}\right) \quad (S2)$$
$$=e^{-2iH_0N_3\tau_{rs}}U_{rs}\left(\pm N_3\right)e^{-2iH_0N_2\tau_{pq}}U_{pq}\left(\pm N_2\right)e^{-2iH_0N_1\tau_{mn}}U_{mn}\left(\pm N_1\right).$$

Here we have assumed that both $N_1$ and $N_2$ is even. There are three other cases: (i) For odd $N_1$ and even $N_2$, $U_{pq}(\pm N_2)$ and $U_{rs}(\pm N_3)$ in the above formula should be replaced by $U_{pq}(\mp N_2)$ and $U_{rs}(\mp N_3)$; (ii) For even $N_1$ and odd $N_2$, $U_{rs}(\pm N_3)$ should be replaced by $U_{rs}(\mp N_3)$; (iii) For odd $N_1$ and odd $N_2$, $U_{pq}(\pm N_2)$ should be replaced by $U_{pq}(\mp N_2)$. For all these cases, we can still use this formula in Eq. (S2) but modify the phases of $\beta_{pq}$ and $\beta_{rs}$ correspondingly. Therefore, the sensor coherence dip is

$$L^{\text{dip}}(N_1,N_2,N_3)=\frac{1}{d}\text{Tr}\left[\left(U^{(-)}_{(N_1,N_2,N_3)}\right)^\dagger U^{(+)}_{(N_1,N_2,N_3)}\right]$$
$$\approx \frac{1}{d}\text{Tr}\left[U_{mn}(N_1)e^{2iH_0N_1\tau_{mn}}U_{pq}(N_2)e^{2iH_0N_2\tau_{pq}}U_{rs}(2N_3)e^{-2iH_0N_2\tau_{pq}}U_{pq}(N_2)e^{-2iH_0N_1\tau_{mn}}U_{mn}(N_1)\right] \quad (S3)$$
$$=\frac{1}{d}\text{Tr}\left[U_{rs}(2N_3)e^{-2iH_0N_2\tau_{pq}}U_{pq}(N_2)e^{-2iH_0N_1\tau_{mn}}U_{mn}(2N_1)e^{2iH_0N_1\tau_{mn}}U_{pq}(N_2)e^{2iH_0N_2\tau_{pq}}\right]$$
$$=\frac{1}{d}\text{Tr}\left[U_{rs}(2N_3)e^{-2iH_0N_2\tau_{pq}}U_{pq}(N_2)e^{2iH_0N_2\tau_{pq}}e^{-iH_0(2N_1\tau_{mn}+2N_2\tau_{pq})}U_{mn}(2N_1)e^{iH_0(2N_1\tau_{mn}+2N_2\tau_{pq})}e^{-2iH_0N_2\tau_{pq}}U_{pq}(N_2)e^{2iH_0N_2\tau_{pq}}\right].$$

The free evolution operators $e^{\pm 2iH_0N_1\tau_{mn}}$ and $e^{\pm 2iH_0N_2\tau_{pq}}$ only change the sign of off-diagonal matrix elements of $U_{pq}(N_2)$ when the pulse numbers $N_2$ is odd, and add a phase factor $(-1)^{N_1}e^{2iN_2\omega_{mn}\tau_{pq}}$ to the off-diagonal matrix elements of $U_{mn}(2N_1)$ (equivalent to modifying $\beta_{mn}$ by multiplying the same phase factor), and therefore will not affect the sensor coherence (since the sensor coherence is generally real and only depends on the modulus of $\beta_{mn}$, except that the sensor coherence for the ring-type transition in Fig. S4(a) has an fast-oscillating imaginary part which truly depends on the argument of $\beta_{mn}$, in this case we only consider the real part of the sensor coherence in Eq. (S6) below which only depends on the modulus of $\beta_{mn}$). Then the sensor coherence dip can be simplified as

$$L^{\text{dip}}(N_1, N_2, N_3) \approx \frac{1}{d}\text{Tr}\left[U_{rs}(2N_3)U_{pq}(N_2)U_{mn}(2N_1)U_{pq}(N_2)\right], \tag{S4}$$

The sensor coherence as a function of $N_1, N_2, N_3$ has different forms depending on the specific correlation types of the three transitions.

(a) *Uncorrelated transitions* - the three transitions share no common state ($d \geq 6$),

$$L^{\text{dip}}_{\text{Uncorre}}(N_1, N_2, N_3) \approx \frac{1}{d}\left[d - 6 + 2\cos(2N_1\delta_1) + 2\cos(2N_2\delta_2) + 2\cos(2N_3\delta_3)\right]. \tag{S5}$$

(b) *Ring-type correlation* - the three transitions are among three states [$d \geq 3$ and type-$\Delta$ in Fig. S4(a)],

$$\begin{aligned}L^{\text{dip}}_{\Delta}(N_1, N_2, N_3) \approx \frac{1}{d}\Big[&d - 3 + \cos(2N_1\delta_1)\cos(2N_3\delta_3) - \sin^2(N_2\delta_2) \\&+ \cos(2N_1\delta_1)\cos^2(N_2\delta_2) + \cos^2(N_2\delta_2)\cos(2N_3\delta_3) \\&- \cos(2N_1\delta_1)\sin^2(N_2\delta_2)\cos(2N_3\delta_3)\Big].\end{aligned} \tag{S6}$$

(c) *Star-type correlation* - the three transitions share one state ($d \geq 4$). The sensor coherence has the same form as that for the ring-type correlation.

(d) *Linked ladder-type correlations* - the three transitions form a linked ladder [$d \geq 4$ and ladder type (*linked*) in Fig. S4(a)],

$$\begin{aligned}L^{\text{dip}}_{\text{ladder}(linked)}(N_1, N_2, N_3) \approx \frac{1}{d}\Big[&d - 4 + \cos(2N_1\delta_1) - \sin^2(N_2\delta_2) + \cos(2N_3\delta_3) \\&+ \cos(2N_1\delta_1)\cos^2(N_2\delta_2) + \cos^2(N_2\delta_2)\cos(2N_3\delta_3) \\&- \cos(2N_1\delta_1)\sin^2(N_2\delta_2)\cos(2N_3\delta_3)\Big].\end{aligned} \tag{S7}$$

(e) *Unlinked ladder-type correlations* - the three transitions form an unlinked ladder type transition [$d \geq 5$ and ladder type (*unlinked*) in Fig. S4(a)],

$$\begin{aligned}L^{\text{dip}}_{\text{ladder}(unlinked)}(N_1, N_2, N_3) \approx \frac{1}{d}\Big[&d - 5 + 2\cos(2N_1\delta_1) + \cos(2N_2\delta_2) + \cos(2N_3\delta_3) \\&+ \cos(2N_2\delta_2)\cos(2N_3\delta_3)\Big].\end{aligned} \tag{S8}$$

The sensor coherence for both correlated and uncorrelated transitions is periodic with the unit cell $N_{1c} \times N_{2c} \times N_{3c}$ ($N_{ic} = \pi/\delta_i$) in the 3D space $(N_1, N_2, N_3)$. Note that if one exchanges $N_1, \delta_1$ and $N_3, \delta_3$ the sensor coherence remains unchanged due to the symmetry in Eq. (S4). In Fig. S4(b) we show the sensor coherence as a function of $N_1, N_2$ while keeping $N_3$ constant for three types of correlations (b), (d), and (e), and find that the coherence patterns are different for different correlation types.